\documentclass[12pt]{iopart}
\usepackage{graphicx}

\begin{document}

\title{Determining layer number of two dimensional flakes of transition-metal dichalcogenides by the Raman intensity from substrate}

\author{Xiao-Li Li,  Xiao-Fen Qiao,  Wen-Peng Han,  Xin Zhang,  Qing-Hai Tan,  Tao Chen,  Ping-Heng Tan}

\address{State Key Laboratory of Superlattices and Microstructures, Institute of Semiconductors, Chinese Academy of Sciences, Beijing 100083, China}
\ead{phtan@semi.ac.cn}
\vspace{10pt}
%\begin{indented}
%\item[]February 2014
%\end{indented}

\begin{abstract}
Transition-metal dichalcogenide (TMD) semiconductors have been widely studied due to their distinctive electronic and optical properties. The property of TMD flakes is a function of its thickness, or layer number (N). How to determine N of ultrathin TMDs materials is of primary importance for fundamental study and practical applications. Raman mode intensity from substrates has been used to identify N of intrinsic and defective multilayer graphenes up to N=100. However, such analysis is not applicable for ultrathin TMD flakes due to the lack of a unified complex refractive index ($\tilde{n}$) from monolayer to bulk TMDs. Here, we discuss the N identification of TMD flakes on the SiO$_2$/Si substrate by the intensity ratio between the Si peak from 100-nm (or 89-nm) SiO$_2$/Si substrates underneath TMD flakes and that from bare SiO$_2$/Si substrates. We assume the real part of $\tilde{n}$ of TMD flakes as that of monolayer TMD and treat the imaginary part of $\tilde{n}$ as a fitting parameter to fit the experimental intensity ratio. An empirical $\tilde{n}$, namely, $\tilde{n}_{eff}$, of ultrathin MoS$_{2}$, WS$_{2}$ and WSe$_{2}$ flakes from monolayer to multilayer is obtained for typical laser excitations (2.54 eV, 2.34 eV, or 2.09 eV). The fitted $\tilde{n}_{eff}$ of MoS$_{2}$ has been used to identify N of MoS$_{2}$ flakes deposited on 302-nm SiO$_2$/Si substrate, which agrees well with that determined from their shear and layer-breathing modes. This technique by measuring Raman intensity from the substrate can be extended to identify N of ultrathin 2D flakes with N-dependent $\tilde{n}$ . For the application purpose, the intensity ratio excited by specific laser excitations has been provided for MoS$_{2}$, WS$_{2}$ and WSe$_{2}$ flakes and multilayer graphene flakes deposited on Si substrates covered by 80-110 nm or 280-310 nm SiO$_2$ layer.

\end{abstract}

%\pacs{75.47.-m, 63.22.Np, 81.15.Cd}% insert suggested PACS numbers in braces on next line

\maketitle %\maketitle must follow title, authors, abstract and \pacs
\section{Introduction}
With the advent of graphene and the exfoliation technique for preparing atomically thin sheets,\cite{Novoselov-Science-2004} layered materials(LMs) sparked wide interest in the world.\cite{Chhowalla-natchem-2013} Among the LMs, transition-metal dichalcogenide(TMD) 2H-MX$_2$(M=Mo,W;X=S,Se) semiconductors have been widely studied due to their distinctive electronic and optical properties.\cite{mak-prl-2010,
Wang-natnano-2012,zhang-ChemSocRev-15} They have an X-M-X covalently bonded sandwich structure in each layer, and the layers are weakly stacked by van der Waals force. Such stacked layer structure makes it possible to peel off different layers from bulk. The property of TMD flakes is a function of its thickness, or layer number (denoted as N).\cite{mak-prl-2010,Splendiani-nanolett-2010,Lee-acsnano-2010,zhang-PRB-13,Terrones-SciRep-2014,Lee-Nanoscale-15}. For example, the band gap of MoS$_2$, WS$_2$ and WSe$_2$ exhibits an indirect-to-direct transition from a few-layer to monolayer thickness,\cite{mak-prl-2010,Splendiani-nanolett-2010} enabling many applications in electronics and optoelectronics. Thus, how to determine N of ultrathin TMDs materials is of primary importance for fundamental study and practical applications.

Several optical techniques have been developed to identify N of the TMD flakes, such as photoluminescence (PL) and optical contrast.\cite{zhang-ChemSocRev-15} PL can be used to distinguish 1L from multilayer (ML) because of its strong and narrow PL peak.\cite{mak-prl-2010} Optical contrast is not sensitive to the flake quality and the stacking structure of 2D flakes,\cite{Ni-nl-07,Casiraghi-nl-07,Han-aps-13} and thus it has been widely used to identify N of graphene flakes by comparing the experimental value with the theoretical one for different N thanks to almost identical complex refractive index ($\tilde{n}$) from 1L graphene (1LG) to ML graphene (MLG).\cite{Han-aps-13,Lu-SB-15} However, quantitative analysis of optical contrast of ultrathin TMD flakes on SiO$_2$/Si substrate is difficult because there exist abundant features associated with optical transitions in the wavelength ($\lambda$) dependent $\tilde{n}$ for TMD flakes and $\tilde{n}$ of TMD flakes itself significantly depends on N due to the indirect-to-direct transition from ML to 1L.\cite{Heinz-PRB-14,Park-JAP-2014}

The ultra-low Raman spectroscopy has been used to reliably determine N for MLG, TMD flakes and 2D alloy flakes.\cite{tan-2012-NM-shear,zhang-PRB-13,qxf-APL-15} However, this technique requires expensive adapters and nonstandard equipment setup. Therefore, it is essential to look for the technique for N identification only by the standard Raman system. Recently, Raman mode intensity from substrates has been used to identify N of intrinsic and defective MLGs up to $N=$100.\cite{Li-Nanoscale-15} This technique is difficult to be applied to the TMD flakes due to the lake of unified $\tilde{n}$ for TMD flakes from 1L to ML. However, the calculation of Raman mode intensity from substrates only requires $\tilde{n}$ at the wavelengths of the laser excitation and the scattered photon, and thus, in this letter, we try to extend this technique for N determination of TMD flakes deposited onto the SiO$_2$/Si substrate. By fitting the experimental data of the intensity ratio between the Si peak from SiO$_2$/Si substrates underneath TMD flakes and that from bare SiO$_2$/Si substrates, we obtained empirical $\tilde{n}$, namely, $\tilde{n}_{eff}$, for ultrathin MoS$_{2}$, WS$_{2}$ and WSe$_{2}$ flakes at different laser excitation wavelengths. The fitted $\tilde{n}_{eff}$ of MoS$_{2}$ has been used to determine N of MoS$_{2}$ flakes on Si substrate covered by 302 nm SiO$_2$, which agrees well with that determined from their shear and layer-breathing modes\cite{zhang-PRB-13}.

\section{Experimental details}
Ultra-thin MoS$_{2}$, WS$_{2}$ and WSe$_{2}$ flakes were mechanically exfoliated from bulk MoS$_{2}$, WS$_{2}$ and WSe$_{2}$ (purchased from 2d Semiconductors, Inc.), and transferred onto Si substrates covered with SiO$_2$ film with different thickness ($h_{SiO_2}$, 100 nm, 89 nm, or 302 nm). Raman measurements were performed at room temperature using a Jobin-Yvon HR800 micro-Raman system equipped with a liquid nitrogen-cooled charge couple detector (CCD), a $\times$50 objective lens with
a numerical aperture (N.A.) of $\sim$0.45 and a 1800 lines/mm grating. The excitation energies ($\varepsilon_L$) are 2.09 eV from a He-Ne laser, 2.34 eV and 2.54eV from a Kr$^+$ laser. Plasma lines were removed from the laser beam by BragGrate Bandpass filters. Measurements down to 5 cm$^{-1}$ are enabled by three BragGrate notch filters with optical density 3 and with full width at half maximum (FWHM) of 8cm$^{-1}$.\cite{zhang-PRB-13} Both BragGrate bandpass and notch filters are produced by OptiGrate Corp. The typical laser power is about 0.4 mW to avoid sample heating.

\section{Results and discussion}
\subsection{$\tilde{n}_{eff}$ of MoS$_{2}$ flakes fitted from N-dependent Si mode intensity}

\begin{figure*}[tb]
\centerline{\includegraphics[width=140mm,clip]{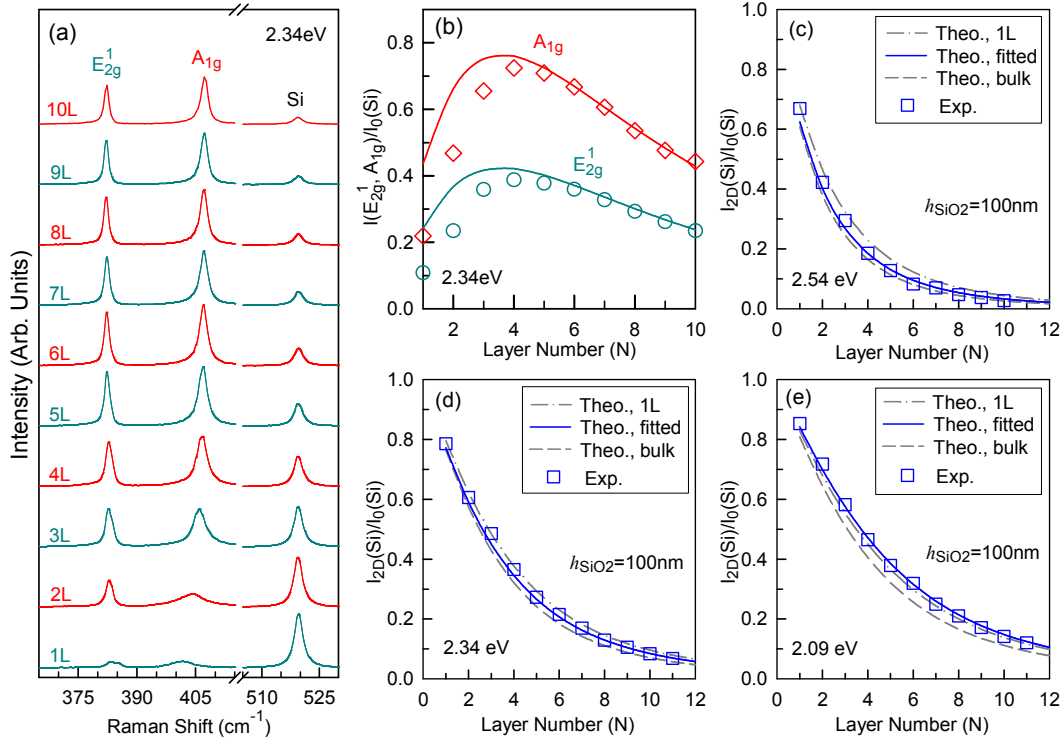}}
\caption{(a) Raman spectra of NL-MoS$_2$ with N from 1 to 10 in the range of the E$^1_{2g}$, A$_{1g}$ and Si modes for $\varepsilon_L$=2.34 eV, where N is determined by the C and LB modes of the flakes. (b) N-dependent I(E$^1_{2g}$)/I$_0$(Si) and I(A$_{1g}$)/I$_0$(Si) and the corresponding theoretical curves calculated from $\tilde{n}_{eff}$. The experimental (Exp., squares) and theoretical (Theo., solid, dashed, and dash-dotted lines) data of I$_{2D}$(Si)/I$_0$(Si) related with NL-MoS$_{2}$ flakes as a function of N for the excitation energies of (c) 2.54 eV, (d) 2.34 eV and (e) 2.09 eV. $h_{SiO_2}$ = 100 nm. The dashed, dash-dotted and solid lines are calculated based on the $\tilde{n}$ for 1L-MoS$_{2}$ and bulk-MoS$_{2}$ and the fitted $\tilde{n}_{eff}$ for NL-MoS$_{2}$, respectively.} \label{Fig1}
\end{figure*}

We denote a N-layer TMD flake as NL-TMD, such as NL-MoS$_2$, NL-WS$_2$ and NL-WSe$_2$, and thus monolayer MoS$_2$ is denoted as 1L-MoS$_2$. 1-10L MoS$_{2}$ flakes were pre-estimated by the Raman measurements of the ultralow-frequency shear (C) and layer-breathing (LB) modes, as previously done for MoS$_{2}$ and MoWS$_2$ flakes.\cite{zhang-PRB-13,qxf-APL-15} Fig.\ref{Fig1}(a) shows the high-frequency Raman spectra of 1L-10L MoS$_{2}$ along with Si mode from substrate were measured by $\varepsilon_L$ of 2.34 eV. Because of the different symmetry between even and odd NL-TMDs and bulk TMDs, the corresponding two Raman-active modes E$^1_{2g}$ and A$_{1g}$ in bulk TMDs should be assigned as the E$'$ and A$_1'$ in odd NL-TMDs and the E$_g$ and A$_{1g}$ modes in even NL-TMDs,\cite{zhang-ChemSocRev-15,Molina-prb-11} respectively. However, to see the evolution from 1L to NL ($N>$1), hereafter the two modes for all cases are simply labeled as E$^1_{2g}$ and A$_{1g}$, as commonly done in the literature.\cite{zhang-ChemSocRev-15,Lee-acsnano-2010,Li-ACSNano-2012} With increasing N, the peak position difference between the E$^1_{2g}$ and A$_{1g}$ modes, $\Delta\omega(A-E)$=Pos(A$_{1g}$)-Pos(E$^1_{2g}$), increases from 17.4 cm$^{-1}$ for 1L to 25 cm$^{-1}$ for 10L, following the formula of $\Delta\omega(A-E)=25.8-8.4/N$.\cite{zhang-ChemSocRev-15} The peak area of the E$^1_{2g}$ and A$_{1g}$ modes, I(E$^1_{2g}$) and I(A$_{1g}$), increases with N up to N=4 and then gradually decreases with N, as shown in Fig.\ref{Fig1}(b), after normalized by the peak area of the Si mode (I$_0$(Si)) from bare substrate that is not covered by MoS$_2$ flakes. Obviously, I(E$^1_{2g}$)/I$_0$(Si) and I(A$_{1g}$)/I$_0$(Si) can not be used to identify N for few-layer MoS$_2$.

Fig.\ref{Fig1}(a) shows that peak area of the Si mode, I$_{2D}$(Si), from the substrate underneath MoS$_2$ flakes monotonously decreases with increasing N. Now we focus on I$_{2D}$(Si) itself. To exclude the effect of crystal orientation on the Raman intensity, I$_{2D}$(Si) is normalized by I$_0$(Si). The maximum of I$_0$(Si) can be obtained by rotating the Si wafer and adjusting the focus of laser beam onto the Si substrate, then, we directly moved the laser spot to MoS$_{2}$ flake to measure I$_{2D}$(Si) to ensure a good signal-to-noise ratio of I$_{2D}$(Si)/I$_0$(Si). Figs.\ref{Fig1}(c-e) depicts the N-dependent I$_{2D}$(Si)/I$_0$(Si) (squares) for NL-MoS$_{2}$ flakes on SiO$_2$/Si substrate ($h_{SiO_2}$ = 100 nm) excited by $\varepsilon_L$ of 2.54 eV, 2.34 eV and 2.09 eV, clearly showing the monotonous decrease in intensity with increasing N. The laser beam to Si substrate is initially adsorbed by the MoS$_{2}$ flake and the Raman signal from Si substrate is adsorbed again by the MoS$_{2}$ flake, similar to the case of MLGs.\cite{Li-Nanoscale-15}. This make I$_{2D}$(Si)/I$_0$(Si) sensitive to N of MoS$_2$ flakes, implying its possibility of N identification for MoS$_2$ flakes.

I$_{2D}$(Si) can be calculated by using multiple reflection interference method and transfer matrix formalism for multilayered structures\cite{Li-Nanoscale-15}, which can be expressed in air/NL-MoS$_2$/SiO$_{2}$/Si four-layer structure as the following equation:
\begin{equation}
\begin{array}{l}
I_{2D}(Si)\propto\int_{0}^{h_{Si}}\int_{0}^{\arcsin(N.A.)}\int_{0}^{2\pi}\int_{0}^{\arcsin(N.A.)}\int_{0}^{2\pi}\\
\sum_{i=s,p_\perp,p_\parallel}\sum_{j=s',p'_\perp,p'_\parallel}\left|F_{L}^{i}(z,\theta,\varphi)F_{R}^{j}(z,\theta',\varphi')\right|^{2}
{\sin\theta\cos\theta}d\theta{d\varphi}\sin\theta'\cos\theta'd\theta'{d\varphi'}dz,
\end{array}
\end{equation}
\noindent where the Raman intensity is given by integrating over the solid angle ($\arcsin(N.A.)$) of microscope objective ($\theta,\varphi$ for the laser beam and $\theta',\varphi'$ for the Raman signal) and the penetration depth of laser excitation into Si layer ($h_{Si}\approx2\mu{m}$). Different from the case of normal incidence where $\theta,\varphi,\theta',\varphi'$=0, the s-polarization (transverse electric field, $\vec{E}$, perpendicular to the NL-MoS$_2$ c-axis) and the p-polarization (transverse magnetic field, $\vec{H}$, associated to electric field by $\vec{H}=\tilde{n}\vec{E}$) field components are considered separately for the oblique incidence, which are involved in the laser excitation enhancement factor $F_{L}$ and Raman scattering enhancement factor $F_{R}$. $F_{L}$ and $F_{R}$ are calculated by using transfer matrix formalism, in which complex refractive index ($\tilde{n}$) and thickness ($h$) of each medium should be known in advance. I$_0$(Si) can be obtained by setting the thickness of MoS$_2$ flakes to be zero. The detailed derivation process can be obtained in supplementary data.

I$_{2D}$(Si)/I$_0$(Si) is expected to be sensitive to N, N.A. of the objective used, $\varepsilon_L$ and $h_{SiO_2}$, as demonstrated in the case of MLGs.\cite{Li-Nanoscale-15} A variation of 10 nm for $h_{SiO_2}$ can introduce a change on the N-dependent I$_{2D}$(Si)/I$_0$(Si),\cite{Li-Nanoscale-15} therefore, a precise determination of $h_{SiO_2}$ is very important for N identification of NL-MoS$_2$ flakes on SiO$_2$/Si substrates by I$_{2D}$(Si)/I$_0$(Si). As a simple, fast and nondestructive technique, optical contrast measurement can be used to determine $h_{SiO_2}$ with a typical micro-Raman confocal system.\cite{Lu-SB-15} It is found that an effective N.A. must be used to calculate optical contrast of multilayer graphene deposited on SiO$_2$/Si substrates once the commonly-used 100$\times$ objective with N.A. of $\sim$ 0.9 is used for optical microscope.\cite{Casiraghi-nl-07,Han-aps-13,Li-Nanoscale-15} In this work, the 50$\times$ objective with N.A. of 0.45 is used to measure I$_{2D}$(Si)/I$_0$(Si). In fact, as shown in the Supplementary Data, it is found that I$_{2D}$(Si)/I$_0$(Si) is not sensitive to N.A. when N.A.$\leq$ 0.5 and N $\leq$ 10 for NL-MoS$_2$ flakes on SiO$_2$/Si substrates.

The unified $\tilde{n}$=$n$-$i$$\kappa$ for MoS$_2$ flakes from 1L to ML is necessary to calculate N-dependent I$_{2D}$(Si)/I$_0$(Si), where $n$ and $\kappa$ are the real and imaginary parts of $\tilde{n}$, respectively. However, both $n$ and $\kappa$ for MoS$_2$ flakes are found to be sensitive to N in the visible region.\cite{Heinz-PRB-14,Park-JAP-2014} $\lambda$-dependent $\tilde{n}$ of 1L-MoS$_{2}$ and bulk MoS$_{2}$ are obtained according to their complex dielectric functions\cite{Heinz-PRB-14} using a formula of $\tilde{n}^{2}=\tilde{\varepsilon}$. If we apply $\tilde{n}$ of 1L-MoS$_{2}$ or bulk MoS$_{2}$ to all the NL-MoS$_{2}$ flakes, I$_{2D}$(Si)/I$_0$(Si) can be calculated for the three $\varepsilon_L$, as shown in Figs.\ref{Fig1}(c-e) by dash-dotted and dashed lines, respectively. Both of them do not fit well to the experimental data. When $\varepsilon_L$=2.54 eV and 2.34 eV, the experimental data lay in between the two theoretical curves. However, when $\varepsilon_L$=2.09 eV excitation, the experimental data are larger than the two theoretical ones because it is under the near-resonant condition with the B exciton.\cite{mak-prl-2010,Lee-Nanoscale-15}

In order to identify N by I$_{2D}$(Si)/I$_0$(Si), an empirical $\tilde{n}$, namely, $\tilde{n}_{eff}$, is necessary to be adopted for MoS$_{2}$ flakes to minimize the difference between the theoretical and experimental data. If we do not consider the multiple reflection interference effect, the difference between I$_{2D}$(Si) and I$_0$(Si) results from the adsorption of the laser beam and Raman beam when they pass through the MoS$_{2}$ flakes, which is mainly dominated by the imaginary part ($\kappa$) of $\tilde{n}$ of MoS$_{2}$, but not by the real part ($n$) of $\tilde{n}$ of MoS$_{2}$. Thus, we assume the real part of $\tilde{n}_{eff}$, namely, $n_{eff}$, of MoS$_{2}$ flakes as $n$ of 1L-MoS$_{2}$. Also, as an approximation, we neglect the difference of $\tilde{n}_{eff}$ between the wavelengths of laser excitation and Raman beam. Finally, we can obtain the imaginary part of $\tilde{n}_{eff}$, namely, $\kappa_{eff}$, of MoS$_{2}$ flakes by fitting the experimental I$_{2D}$(Si)/I$_0$(Si) by the theoretical ones for each excitation wavelength. We found that, indeed, a $\kappa_{eff}$ can make the theoretical I$_{2D}$(Si)/I$_0$(Si) agree well with the experimental ones, as shown in Figs.\ref{Fig1}(c-e) by solid curves. The fitted $\kappa_{eff}$ along with $n_{eff}$ for MoS$_{2}$ flakes are summarized in Table 1 for $\varepsilon_L$ of 2.54 eV, 2.34 eV and 2.09 eV. In this case, we can calculate I$_{2D}$(Si)/I$_0$(Si) based on the fitted $\tilde{n}_{eff}$ for MoS$_{2}$ flakes and compare them with the experimental one, N of MoS$_{2}$ flakes can be determined. Based on the fitted $\tilde{n}_{eff}$ for MoS$_{2}$ flakes, we calculated I(E$^1_{2g}$)/I$_0$(Si) and I(A$_{1g}$)/I$_0$(Si) as a function of N for $\varepsilon_L$=2.34 eV, where adjustable parameters were introduced to take different efficiencies among E$^1_{2g}$, A$_{1g}$ and Si modes into account. The results are shown by solid corves in Fig.\ref{Fig1}(b). The theoretical results basically correspond with the experimental ones, but there exists significant discrepancy for $N<5$.

\subsection{$\tilde{n}_{eff}$ of NL-WS$_2$ and NL-WSe$_2$ flakes}

\begin{figure*}[tb]
\centerline{\includegraphics[width=140mm,clip]{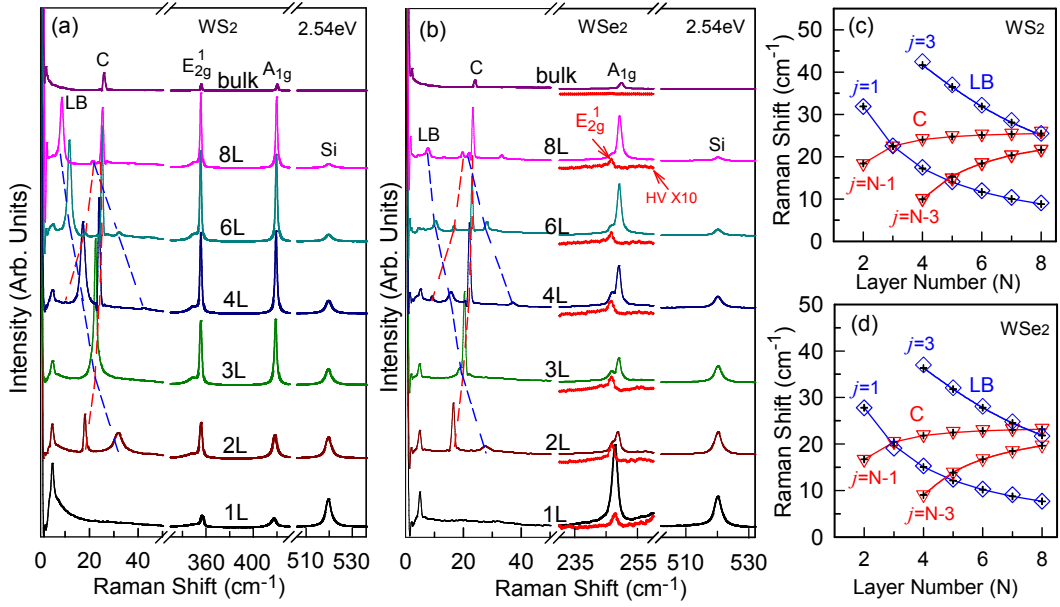}}
\caption{Raman spectra of (a) NL-WS$_2$ and (b) NL-WSe$_2$ in the ultra-low frequency range and in the range of the E$^1_{2g}$, A$_{1g}$ and Si modes for $\varepsilon_L$=2.54 eV. Pos(C) and Pos(LB) in (c) NL-WS$_2$ and (d) NL-WSe$_2$ as a function of N. $j$=1,3 and $j$=N-1 and N-3 correspond to the observed branches of the Sin diagram, see text. } \label{Fig2}
\end{figure*}

Now we check the possibility to apply this technique to other TMDs, such as WS$_{2}$ and WSe$_{2}$. We obtain 1-8L WS$_{2}$ and WSe$_{2}$ flakes by mechanical exfoliation from bulk crystal and N is determined by the C and LB modes based on the method described in Ref.\cite{qxf-APL-15}. The C and LB modes and the E$^1_{2g}$ and A$_{1g}$ modes of 1-8L WS$_{2}$ and WSe$_{2}$ were measured by 2.54-eV laser excitation at room temperature, as depicted in Figs.\ref{Fig2}(a) and \ref{Fig2}(b), respectively. Pos(C) and Pos(LB) of 2-8L WS$_{2}$ and WSe$_{2}$ are summarized in Figs.\ref{Fig2}(c) and \ref{Fig2}(d), respectively. Besides a fan diagram\cite{zhang-PRB-13}, the N-dependent Pos(C) and Pos(LB) can also exhibit a Sin diagram\cite{Wu-acsnano-2015}, which can be written as follows:
\begin{equation}
\eqalign{\omega(C_{N,N-j})=\sqrt{2}\omega(C_{21})sin(\frac{j\pi}{2N}),\cr
\omega(LB_{N,N-j})=\sqrt{2}\omega(LB_{21})sin(\frac{j\pi}{2N}).}
\end{equation}
\noindent where $j$ is an integer, $j$=N-1,N-2,...,2,1. $\omega(C_{21})$ and $\omega(LB_{21})$ are the frequencies of the C and LB modes in 2L flakes, respectively. The measured C modes in bulk WS$_{2}$ and WSe$_{2}$ are located at 26.3 and 23.9 cm$^{-1}$, and then the measured $\omega(C_{21})$ ($\omega(LB_{21})$) of 2L-WS$_{2}$ and 2L-WSe$_{2}$ are 18.4 (31.9) and 16.7 (27.8) cm$^{-1}$, respectively. Each branch in Eq.~(1) always decreases or increases in frequency with increasing N. As indicted by the solid lines in Figs.\ref{Fig2}(c) and \ref{Fig2}(d), the branches of $j=N-1$ and $j=N-3$ are observed for the C modes, and the branches of $j=1$ and $j=3$ are observed for the LB modes.

The E$^1_{2g}$ and A$_{1g}$ modes of 1-8L WS$_{2}$ and WSe$_{2}$ are found to be insensitive to N, as addressed in previous works\cite{Terrones-SciRep-2014,zhao-Nanoscale-13}. The E$^1_{2g}$ mode of WSe$_{2}$ flakes is very weak and it can be clearly revealed under cross (HV) polarization configuration, as shown in Fig.\ref{Fig2}(b). The frequency shift of both E$^1_{2g}$ and A$_{1g}$ modes for NL-WS$_{2}$ and NL-WSe$_{2}$ is less than 3 cm$^{-1}$, which make it difficult for N determination.

\begin{figure*}[tb]
\centerline{\includegraphics[width=140mm,clip]{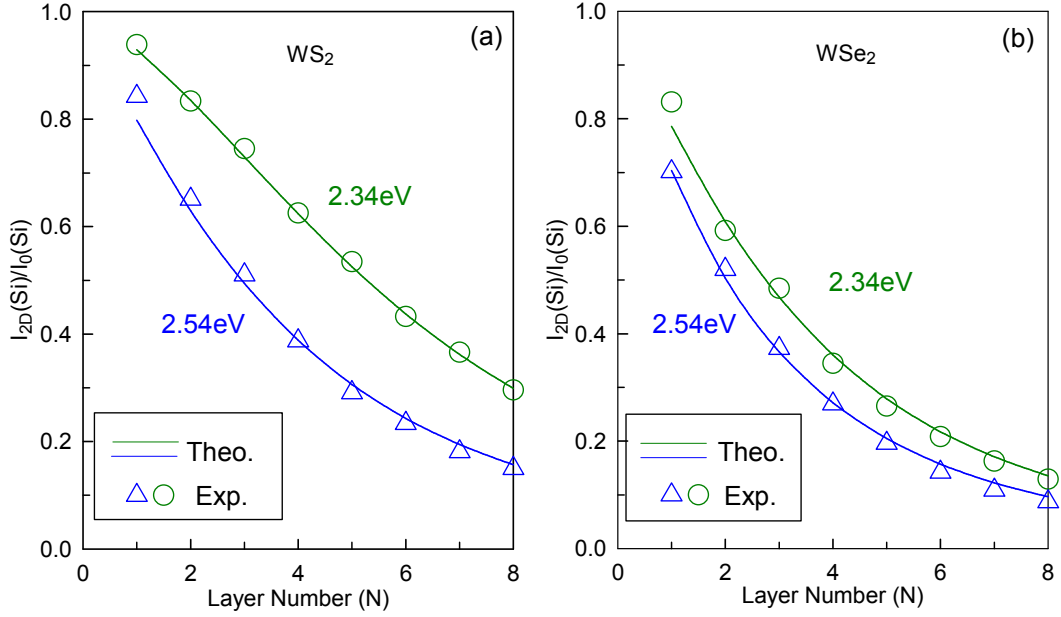}}
\caption{The experimental (Exp., triangles and circles) and theoretical (Theo., solid lines) data of I$_{2D}$(Si)/I$_0$(Si) for the excitation energies of $\varepsilon_L$=2.54 eV and 2.34 eV as a function of N. (a) NL-WS$_{2}$, (b) NL-WSe$_{2}$. $h_{SiO_2}$ = 89 nm.} \label{Fig3}
\end{figure*}

I$_{2D}$(Si) from Si substrate underneath WS$_{2}$ and WSe$_{2}$ flakes decreases with increasing N, as shown in Figs.\ref{Fig2}(a) and \ref{Fig2}(b), similar to the case of MoS$_2$ flakes in Fig.\ref{Fig1}(a). This suggests that I$_{2D}$(Si)/I$_0$(Si) can be used for N identification of WS$_{2}$ and WSe$_{2}$ flakes deposited on SiO$_2$/Si substrate. I$_{2D}$(Si)/I$_0$(Si) for NL-WS$_{2}$ and NL-WSe$_{2}$ flakes as a function of N (N=1,2,...,8) for $h_{SiO_2}$ = 89 nm were measured for two excitations: $\varepsilon_L$=2.54 eV and 2.34 eV. Indeed, I$_{2D}$(Si)/I$_0$(Si) monotonously decreases with increasing N for WS$_{2}$ and WSe$_{2}$ flakes excited by the two excitations. The corresponding experimental data were depicted in Fig.\ref{Fig3} by triangles and circles, respectively. Considering that the 2.09-eV excitation is almost resonant with the A exciton of 1L-WS$_{2}$ and B exciton of 1L-WeS$_{2}$,\cite{zhao-AcsNano-13}, the 2.09-eV laser excitation is not used for WS$_{2}$ and WeS$_{2}$ flakes.

To understand the N-dependent I$_{2D}$(Si)/I$_0$(Si) of WS$_{2}$ and WSe$_{2}$ flakes for their N determination, similar to the case of MoS$_2$ flakes as discussed above, we assume $n_{eff}$ of WS$_{2}$ and WSe$_{2}$ flakes as $n$ of 1L-WS$_{2}$ and 1L-WSe$_{2}$\cite{Heinz-PRB-14}, respectively. $\kappa_{eff}$ of WS$_{2}$ and WSe$_{2}$ flakes have been used as a fitting parameter to fit the experimental I$_{2D}$(Si)/I$_0$(Si). The $\tilde{n}_{eff}$ of WS$_{2}$ and WSe$_{2}$ flakes for $\varepsilon_L$=2.54 and 2.34 eV are summarized in Table 1. The calculated I$_{2D}$(Si)/I$_0$(Si) based on the fitted $\tilde{n}_{eff}$ are shown by solid lines in Fig.\ref{Fig3}, which agree well with the experimental values.

\begin{table}
\caption{The empirical $\tilde{n}_{eff}$ ($n_{eff}$-$i\kappa_{eff}$) of MoS$_{2}$, WS$_{2}$ and WSe$_{2}$ flakes to calculate I$_{2D}$(Si)/I$_0$(Si) at different $\varepsilon_L$.}
\begin{center}
\begin{tabular}{c|c|c|c}
  \hline\hline
  % after \\: \hline or \cline{col1-col2} \cline{col3-col4} ...
  {} & {MoS$_{2}$} & {WS$_{2}$} & {WSe$_{2}$} \\ \hline
   {$\varepsilon_L$(eV)}& 2.54   2.34   2.09 &  2.54   2.34    &  2.54   2.34   \\
   {$n_{eff}$}& 5.29  4.85  4.58 & 4.40  4.62   & 4.22  4.64  \\
   {$k_{eff}$}& 1.85  1.20  1.22 & 1.10  0.48   & 1.86  1.40  \\ \hline\hline
\end{tabular}
\end{center}
\end{table}

\subsection{Identifying Layer number of MoS$_{2}$ flakes deposited on 302-nm SiO$_2$/Si substrate}

\begin{figure*}[tb]
\centerline{\includegraphics[width=140mm,clip]{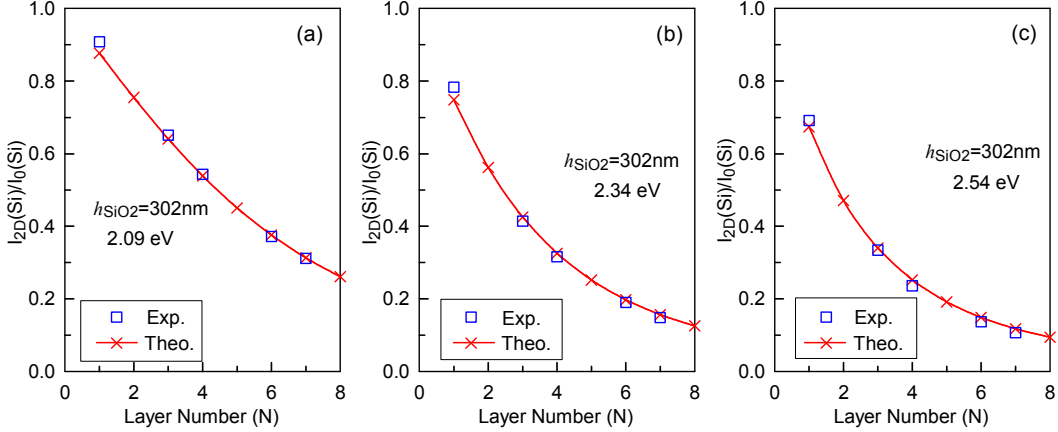}}
\caption{The experimental (Exp., squares) and theoretical (Theo., solid lines and crosses) data of I$_{2D}$(Si)/I$_0$(Si) related with NL-MoS$_{2}$ flakes as a function of N for the excitation energies of (a) 2.09 eV, (b) 2.34 eV and (c) 2.54 eV. The solid lines are calculated based on the fitted $\tilde{n}_{eff}$ from Figure 2. $h_{SiO_2}$ = 302 nm.} \label{Fig4}
\end{figure*}

As discussed above, the empirical $\tilde{n}_{eff}$ of MoS$_{2}$, WS$_{2}$ and WSe$_{2}$ flakes are obtained by fitting the theoretical I$_{2D}$(Si)/I$_0$(Si) to the experimental data excited by different excitation energies. Once $\tilde{n}_{eff}$ for TMD flakes at specific $\varepsilon_L$ is available, one can compare the theoretical I$_{2D}$(Si)/I$_0$(Si) with the corresponding experimental data excited by the same $\varepsilon_L$ to determine N of the TMD flakes. As an example, we apply this technique to MoS$_{2}$ flakes on Si substrate covered by 302-nm SiO$_2$ film, where N is precisely determined by the C and LB modes of MoS$_{2}$ flakes\cite{zhang-PRB-13} and N = 1, 3, 4, 6, 7. Based on the fitted $\tilde{n}_{eff}$ for MoS$_{2}$ flakes, the theoretical I$_{2D}$(Si)/I$_0$(Si) were calculated as a function of N for $\varepsilon_L$=2.09, 2.34 and 2.54 eV, as shown by crosses and solid lines in Figs.\ref{Fig4}(a-c), respectively. They agree well with the experimental data, as depicted by squares in Fig.\ref{Fig4}. The N-dependent I$_{2D}$(Si)/I$_0$(Si) of TMD flakes is found to be sensitive to $\varepsilon_L$, as demonstrated in Figs.\ref{Fig1} and \ref{Fig3}. Thus, the determined N of TMD flakes by an excitation energy can be confirmed from further measurement by another excitation energy, which leads to an accurate N determination of ultrathin TMD flakes by the Raman intensity from substrate.

\section{Conclusions}

In conclusion, a technique to determine N of TMD flakes such as MoS$_{2}$, WS$_{2}$ and WSe$_{2}$ deposited on SiO$_2$/Si substrate has been proposed by measuring I$_{2D}$(Si)/I$_0$(Si), i.e., the intensity ratio of the Si peak from SiO$_2$/Si substrates underneath the 2D flakes of TMDs to that from bare SiO$_2$/Si substrates. The real part of the empirical $\tilde{n}_{eff}$ of TMD flakes is assumed as that of 1L TMD. The $\kappa_{eff}$ is a fitting parameter to the experimental intensity ratio between the Si peak from SiO$_2$/Si substrates underneath TMD flakes and that from bare SiO$_2$/Si substrates. The empirical $\tilde{n}_{eff}$ of MoS$_{2}$, WS$_{2}$ and WSe$_{2}$ flakes for $\varepsilon_L$ of 2.54, 2.34 or 2.09 eV is obtained. The resulted $\tilde{n}_{eff}$ of MoS$_{2}$ flakes has been used to identify N of MoS$_{2}$ flakes deposited on 302-nm SiO$_2$/Si substrate. This opens the possibility to identify N of ultrathin 2D flakes with N-dependent complex refractive index by measuring Raman intensity from the substrate. For the sake of N identification of TMD and MLG flakes for research community, I$_{2D}$(Si)/I$_0$(Si) of TMD flakes deposited on SiO$_2$/Si substrate is enclosed in the supplementary data for commonly-used excitation energies (2.34 eV and 2.54 eV) and SiO$_2$ thickness (80-110nm and 280-310nm), along with the corresponding data for MLG.

\section{Acknowledgments}

We acknowledge support from the National Natural Science Foundation of China, grants 11225421, 11434010, 11474277 and 11504077.

\section{References}
\bibliography{MX2-LXL}
\bibliographystyle{iopart-num}

\end{document}

%% 文本文件开始，这是必须的指令

%=================== Text begin here =============================================

\begin{center}
\LARGE\textbf{Supplementary Data}
\end{center}

\begin{center}
\LARGE Determining layer number of two dimensional flakes
of multilayer graphenes and transition-metal dichalcogenides by the Raman
intensity from substrate   %% 论文题目
\end{center}

\begin{center}
\rm Xiao-Li Li$^{1}$, \ \ Xiao-Fen Qiao$^{1}$, \ \ Wen-Peng Han$^{1}$, \ \ Xin Zhang$^{1}$, \ \ Qing-Hai Tan$^{1}$, \ \ Tao Chen$^{1}$, \ and \ Ping-Heng Tan$^{1}$
\end{center}

\begin{center}
\begin{footnotesize} \sl
{1.State Key Laboratory for Superlattices andMicrostructures, Institute of Semiconductors,
 Chinese Academy of Sciences, Beijing 100083, China
 E-mail: phtan@semi.ac.cn} \\   %%%% 地址 a)
%${}^{\rm c)}$ \\   %%%% 地址 c)
%%% 更多地址依次往下延续
\end{footnotesize}
\end{center}

\vspace*{2mm}

Content

1. Calculating the Si mode intensity from SiO$_2$/Si substrates underneath TMD flakes (I$_{2D}$(Si)) and that (I$_0$(Si)) from bare SiO$_2$/Si substrates.

2. I$_{2D}$(Si)/I$_0$(Si) for NL-TMDs deposited on SiO$_2$/Si substrate.

3. I$_{2D}$(Si)/I$_0$(Si) for N Layer graphenes deposited on SiO$_2$/Si substrate.

\section{Calculating the Si mode intensity from SiO$_2$/Si substrates underneath TMD flakes (I$_{2D}$(Si)) and that (I$_0$(Si)) from bare SiO$_2$/Si substrates}

Raman intensity in multilayer structure is determined by multiple reflection at the interfaces and optical interference within the medium. We adopt the multiple reflection
interference method\cite{Wang-apl-08,Yoon-prb-09,Koh-acsnano-11,Li-ascnano-12,Li-Nanoscale-15} to calculate the Raman mode intensity of a medium in the multilayer structure. When N layer (NL) transition-metal dichalcogenide (TMD) flakes (NL-MX$_2$) are deposited on SiO$_2$/Si substrate, the four-layer structure can be established, containing air($\tilde{n}_{0}$),
NL($\tilde{n}_{1}$, $d_{1}$), $SiO_{2}$($\tilde{n}_{2}$, $d_{2}$), Si($\tilde{n}_{3}$, $d_{3}$), where $\tilde{n}_{i}$ and $d_{i}$ ($i$=0,1,2,3) are the complex refractive index and the thickness of each medium, respectively, as demonstrated in Fig. S1.

%\iffalse
\begin{figure}[!htb]
\centerline{\includegraphics[width=140mm,clip]{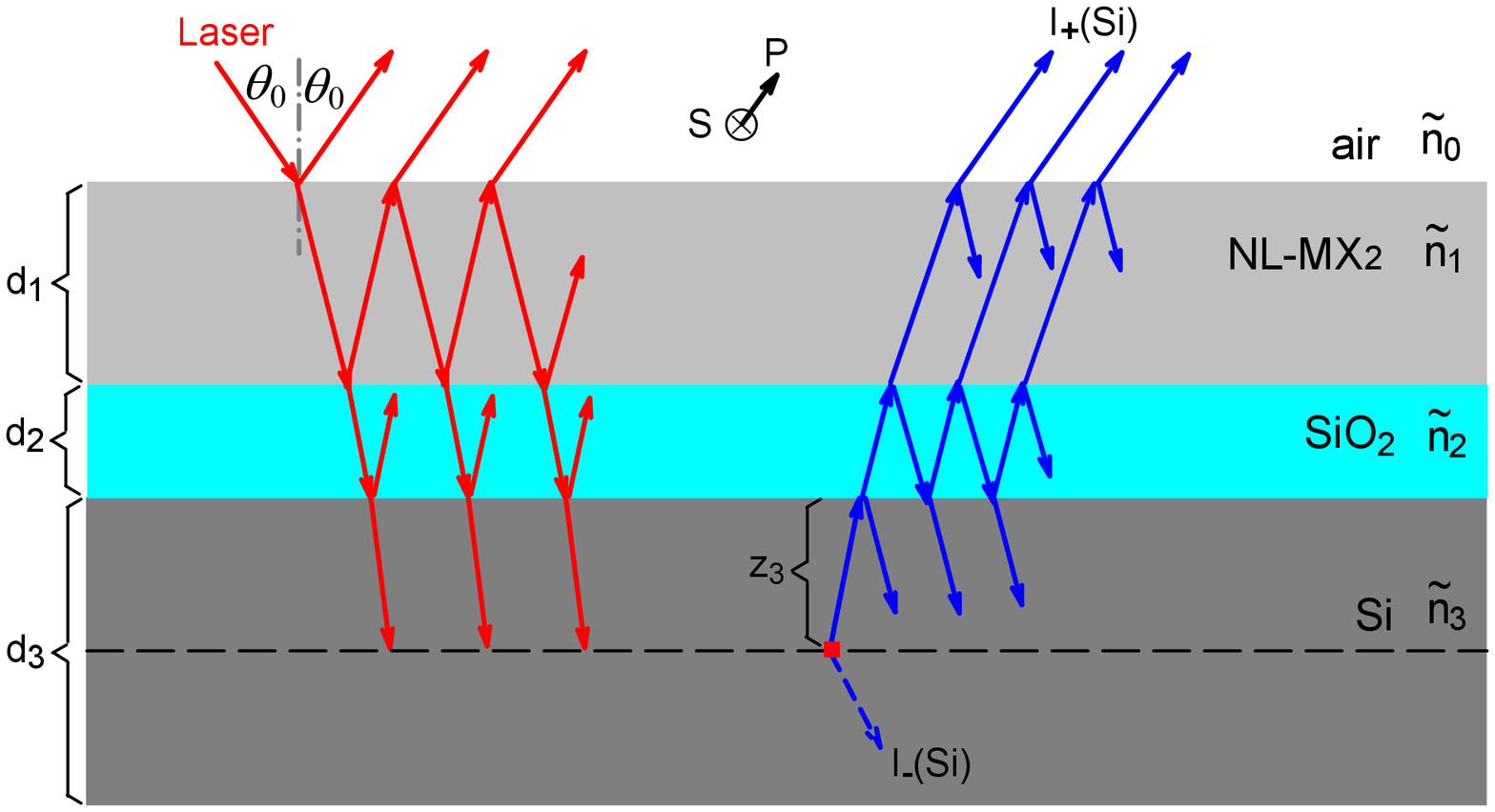}}
\caption{Schematic diagrams of multiple reflection and optical interference in the multilayer structures containing air, NL-MX$_2$, SiO$_2$, and Si for the incident laser and out-going Raman signals of the Si peak from Si substrate. $\tilde{n}_{0}$, $\tilde{n}_{1}$ (d$_1$), $\tilde{n}_{2}$(d$_2$), and $\tilde{n}_{3}$(d$_3$) are the complex refractive indices (thickness) of air, NL-MX$_2$, SiO$_2$ and Si layers, respectively. } \label{FigS1}
\end{figure}
%\fi

In the following text, we will discuss how to calculate the Si mode intensity from SiO$_2$/Si substrates underneath TMD flakes (I$_{2D}$(Si)) and that (I$_0$(Si)) from bare SiO$_2$/Si substrates.

Similar to previous works\cite{Wang-apl-08,Yoon-prb-09,Koh-acsnano-11,Li-ascnano-12,Li-Nanoscale-15}, to calculate the intensity of Raman signal from the multilayer structures, one must treat the laser excitation and Raman scattering processes separately. Raman signals from the depth $z_3$ in the Si layer will be excited by the laser excitation power at the corresponding depth. The multiple reflection and optical interference are also taken into account in the transition process of Raman signal from Si layer to air. We defined $F_{L}$ and $F_{R}$ as respective enhancement factors for laser excitation and Raman signal, similar to the notation of Yoon {\it et al.}\cite{Yoon-prb-09,Li-ascnano-12,Li-Nanoscale-15}. The Raman intensity of Si mode can be expressed by integrating over its thickness, $d_{Si}$, as the following equation:
\begin{equation}
I\propto\int_{0}^{d_{Si}}|F_{L}(z_3)F_{R}(z_3)|^{2}dz_3.
\label{Eq1}
\end{equation}
\noindent The transfer matrix formalism can be used to calculate $F_{L}$ and $F_{R}$ in the multilayer structures, which has been widely used to calculate the Raman signal and optical contrast of NL flakes on SiO$_2$/Si substrate.\cite{Koh-acsnano-11,Casiraghi-nl-07,Han-aps-13,Li-Nanoscale-15} Because the penetration depth of laser excitation into Si layer is far less than the actual thickness of Si substrate ($d_{3}$), the value of $d_{Si}$ is taken as the penetration depth of laser excitation into Si layer in the numerical integration. In order to take the numerical aperture (N.A.) of the objective into account, we calculate contributions from each portion of the laser beam by integrating the incident angle $\theta$ from 0 to $\arcsin(N.A.)$. The s-polarization (transverse electric field, $\vec{E}$, perpendicular to the c-axis) and p-polarization (transverse magnetic field, $\vec{H}$, associated with the electric field by $\vec{H}=\tilde{n}\vec{E}$) field components\cite{Casiraghi-nl-07} are also treated for the transfer matrices. Thus, the total Raman intensity of the Si mode is given by integrating over the solid angles ($\theta,\varphi$ for the laser beam and $\theta',\varphi'$ for the Raman signal) of the microscope objective and the depth $z_3$ in the Si layer:
\begin{equation}
\begin{split}
&I\propto\int_{0}^{d_{Si}}\int_{0}^{\arcsin(N.A.)}\int_{0}^{2\pi}\int_{0}^{\arcsin(N.A.)}\int_{0}^{2\pi}\\
&\sum_{i=s,p_\perp,p_\parallel}\sum_{j=s',p'_\perp,p'_\parallel}\left|F_{L}^{i}(z_3,\theta,\varphi)F_{R}^{j}(z_3,\theta',\varphi')\right|^{2}\\
&{\sin\theta\cos\theta}d\theta{d\varphi}\sin\theta'\cos\theta'd\theta'{d\varphi'}dz_3,
\end{split}
\label{Eq2}
\end{equation}

We use the transfer matrix formalism calculate $F_{L}$ and $F_{R}$ in Eq. (\ref{Eq2}). The transmission and reflection of total electric and magnetic fields in the four-layer structure can be described by characteristic matrices $A_{ij}$ and $B(z_j)$, where $A_{ij}$ describes the propagation across the interface from $i$ to $j$ layer applying the boundary conditions, and $B(z_j)$ denotes the propagation through the $j$ layer at depth $z_j$.

The s-polarization and p-polarization field components are treated separately in analyzing the characteristic matrices. $A^{s(p)}_{ij}$ can be expressed as follows:
\begin{equation}
\begin{split}
A^{s(p)}_{ij}=\frac{1}{t^{s(p)}_{ij}}
\left(
\begin{array}{ccc}
1 & r^{s(p)}_{ij} \\
r^{s(p)}_{ij} & 1
\end{array}
\right)
\end{split}
\label{Eq3}
\end{equation}
\noindent where $t^{s(p)}_{ij}$ and $r^{s(p)}_{ij}$ are transmission and reflection coefficients from $i$ to $j$ layer for s-polarization and p-polarization, which depend on the complex refractive index $\tilde{n}_{i},\tilde{n}_{j}$ and the refractive angle $\theta_i,\theta_j$.
The complex refractive index in the dielectric layer $i$ is denoted by $\tilde{n}_{i}(\lambda)=n_{i}(\lambda)-\textbf{i}k_{i}(\lambda)$, which is dependent on $\lambda$. Similar to previous works,\cite{Li-Nanoscale-15} we consider only the in-plane complex refractive index of NL-MX$_2$ flakes, and thus the s- and p-components of laser excitation can share the same expression. The refracted angle $\theta_i$ in the dielectric layer $i$ is calculated with the Snell law as follows: $t_{ij}^{s}=\frac{2\tilde{n}_i\cos{\theta_i}}{\tilde{n}_i\cos{\theta_i}+\tilde{n}_j\cos{\theta_j}}$, $r_{ij}^{s}=\frac{\tilde{n}_i\cos{\theta_i}-\tilde{n}_j\cos{\theta_j}}{\tilde{n}_i\cos{\theta_i}+\tilde{n}_j\cos{\theta_j}}$, $t_{ij}^{p}=\frac{2\tilde{n}_i\cos{\theta_i}}{\tilde{n}_j\cos{\theta_i}+\tilde{n}_i\cos{\theta_j}}$, $r_{ij}^{p}=\frac{\tilde{n}_j\cos{\theta_i}-\tilde{n}_i\cos{\theta_j}}{\tilde{n}_j\cos{\theta_i}+\tilde{n}_i\cos{\theta_j}}$.

$B(z_j)$ can be expressed as follows:
\begin{equation}
\begin{split}
B(z_j)=
\left(
\begin{array}{ccc}
e^{i\delta(z_j)} & 0 \\
0 & e^{-i\delta(z_j)}
\end{array}
\right)
\end{split}
\label{Eq4}
\end{equation}
\noindent where $\delta(z_j)=2\pi{\tilde{n}_j}z_j\cos{\theta_j}/\lambda$ is phase factor, which is determined by the propagation distance $z_j$ of laser or Raman scattering lights in $j$ layer.
The differences between the wavelengths of the laser and Si Raman peak should be considered in $t_{ij}^{s(p)}$, $r_{ij}^{s(p)}$ and $\delta(z_j)$ due to the $\lambda$-dependent complex refractive index for each layer. Finally, the complete transfer matrix for the whole multilayer structures can be obtained by multiplication of above simple matrices.

In the calculation of I$_{2D}$(Si), the laser approaches at the Si layer and is absorbed finally, and thus there is no laser components transmitting toward the surface at any depths in the Si layer, as illustrated in Fig. S1. The transfer equation for the laser beam transmitting from air to the depth $z_3$ in the Si layer is expressed as follows:
\begin{equation}
\begin{split}
\left(
\begin{array}{ccc}
E_{L0}^{s(p),+} \\
E_{L0}^{s(p),-}
\end{array}
\right)=A^{s(p)}_{01}B(d_1)A^{s(p)}_{12}B(d_2)A^{s(p)}_{23}B(z_3)
\left(
\begin{array}{ccc}
E_{LS}^{s(p),+}(z_3) \\
0
\end{array}
\right)
\end{split}
\label{Eq5}
\end{equation}
\noindent where the symbols $+,-$ denote light propagating directions from air to the dielectric layers and the opposite direction. $E_{L0}^{s(p),+}$ is the electric field component of laser source in air, which is assumed as 1. $E_{LS}^{s(p),+}(z_3)$ is the electric field component at the depth $z_3$ in Si layer. The laser excitation enhancement factors in this process can be calculated by
$F_{L}^{s}(z_3)=E_{LS}^{s,+}(z_3)$, $F_{L}^{p\perp}(z_3)=E_{LS}^{p,+}(z_3)\cos{\theta_3}$ and $F_{L}^{p\parallel}(z_3)=E_{LS}^{p,+}(z_3)\sin{\theta_3}$. $E_{LS}^{s(p),+}(z_3)$ can be calculated with the transfer equation Eq. (\ref{Eq5}) and $F_{L}^{s(p\perp,p\parallel)}(z_3)$ can be obtained.

For the Raman scattering process of the Si mode, only one pathway for the emission of Raman signal toward (up,U) the NL-MX$_2$ surface is considered, and that away from (down,D) the NL-MX$_2$ surface is absorbed by the Si layer, as illustrated in Fig. S1. The transfer equation for the Raman signal excited by the laser at the depth $z_3$ in the Si layer is expressed as follows:
\begin{equation}
\begin{split}
\left(
\begin{array}{ccc}
0\\
E_{R_{U0}}^{s'(p'),-}
\end{array}
\right)=A^{s'(p')}_{01}B(d_1)A^{s'(p')}_{12}B(d_2)A^{s'(p')}_{23}B(z_3)
\left(
\begin{array}{ccc}
E_{R_{US}}^{s'(p'),+}(z_3)\\
E_{R_{US}}^{s'(p'),-}(z_3)
\end{array}
\right)
\end{split}
\label{Eq6}
\end{equation}
\noindent where $E_{R_{US}}^{s'(p'),-}(z_3)$ is the electric field component of Raman signal source related with 'U' pathway at the depth $z_3$ in the Si layer and is assumed as 1.

The Raman scattering enhancement factors in this process can be calculated by $F_{R}^{s'}(z_3)=E_{R_{U0}}^{s',-}$, $F_{R}^{p'_\perp}(z_3)=E_{R_{U0}}^{p',-}\cos{\theta'_3}$
and $F_{R}^{p'_\parallel}(z_3)=E_{R_{U0}}^{p',-}\sin{\theta'_3}$.
$E_{R_{U0}}^{s'(p'),-}$ can be calculated by the transfer equation Eq. (\ref{Eq6}) and $F_{R}^{s'(p'_\perp,p'_\parallel)}(z_3)$ can be obtained.

The total Raman intensity of I$_{2D}$(Si) from SiO$_2$/Si substrates underneath TMD flakes can be calculated by the formula of Eq. (\ref{Eq2}).

I$_0$(Si) from bare SiO$_2$/Si substrates can be calculated directly based on the above model once the thickness of NL flakes is set to zero.

\section{I$_{2D}$(Si)/I$_0$(Si) for NL-TMDs deposited on SiO$_2$/Si substrate}

As demonstrated in the main text, the calculated I$_{2D}$(Si)/I$_0$(Si) of NL-TMDs based on the $\tilde{n}_{eff}$ for different $\varepsilon_L$ can be used for N determination of NL-TMDs flakes deposited on SiO$_2$/Si substrates. I$_{2D}$(Si)/I$_0$(Si) is expected to be sensitive to N, N.A., $\varepsilon_L$ and $h_{SiO_2}$.\cite{Li-Nanoscale-15} It is evident that a variation of 10 nm for $h_{SiO_2}$ can introduce a change on the N-dependent I$_{2D}$(Si)/I$_0$(Si), therefore, a precise determination of $h_{SiO_2}$ is very important for N identification of NL-TMDs flakes on SiO$_2$/Si substrates by I$_{2D}$(Si)/I$_0$(Si).

It is found that an effective N.A. must be used for optical contrast calculation of multilayer graphene deposited on SiO$_2$/Si substrates once the 100$\times$ objective with N.A. of $\sim$ 0.9 is used for optical microscope.\cite{Casiraghi-nl-07,Han-aps-13} If one uses an objective with N.A. less than 0.5 for the measurement of optical contrast, the actual N.A. can be used for the theoretical calculation of optical contrast for N identification of two-dimensional flakes.\cite{Han-aps-13} For the sake of N identification of TMD flakes for research community, I$_{2D}$(Si)/I$_0$(Si) of NL-MoS$_{2}$, NL-WS$_{2}$ and NL-WeS$_{2}$ excited by commonly-used 2.54-eV and 2.34-eV excitations for 80nm$<h_{SiO2}<$110nm and 280nm$<h_{SiO2}<$310nm are calculated and here the objective N.A. of 0.35 (Olympus SLMPLN 50$\times$ objective with a long-working distance of 18mm) and of 0.50 (Olympus LMPLFLN 50$\times$ objective with a long-working distance of 10.6mm) is considered in the calculation. The results are summarized in Table 1-6. We find that I$_{2D}$(Si)/I$_0$(Si) is not sensitive to N.A. when N.A.$\leq$ 0.5 and N $\leq$ 10. 

For an approximation, I$_{2D}$(Si)/I$_0$(Si) for any $h_{SiO2}$ in the range of 80-110nm or 280-310nm can be obtained by the interpolation between the two values corresponding to two nearest $h_{SiO2}$.

\section{I$_{2D}$(Si)/I$_0$(Si) for N Layer graphenes deposited on SiO$_2$/Si substrate}

I$_{2D}$(Si)/I$_0$(Si) has been used to identify N of intrinsic and defective multilayer graphenes up to $N=$100.\cite{Li-Nanoscale-15} For a purpose of practise application, the calculation values of I$_{2D}$(Si)/I$_0$(Si) are provided here for several typical $h_{SiO_2}$ (80-110nm or 280-310nm), commonly-used $\varepsilon_L$ (2.34 and 1.96 eV) and the objective N.A. (0.35 and 0.50), as summarized in Table 7 and Table 8. I$_{2D}$(Si)/I$_0$(Si) is not sensitive to N.A. when N.A.$\leq$ 0.5 and N $\leq$ 15 for NL graphenes deposited on SiO$_2$/Si substrate.

For an approximation, I$_{2D}$(Si)/I$_0$(Si) for any $h_{SiO2}$ in the range of 80-110nm or 280-310nm can be obtained by the interpolation between the two values corresponding to two nearest $h_{SiO2}$.

\begin{sidewaystable}  %Table 1
\caption{I$_{2D}$(Si)/I$_0$(Si) of NL-MoS$_{2}$ flakes deposited on SiO$_2$/Si substrate as a function of N for $h_{SiO2}$=80nm, 90nm, 100nm, 110nm, and 280nm, 290nm, 300nm, 310nm. $\varepsilon_L$=2.54 eV and N.A.=0.35 and 0.50. $\tilde{n}$ of NL-MoS$_{2}$, SiO$_2$ and Si are 5.29-1.85$\bf\emph{i}$, 1.462, 4.360-0.086$\bf\emph{i}$.}
\begin{center}
\begin{tabular}{c|c|c|c|c|c|c|c|c}
  \hline\hline
  % after \\: \hline or \cline{col1-col2} \cline{col3-col4} ...
  {$h_{SiO2}$} & 80nm & 90nm & 100nm & 110nm & 280nm & 290nm & 300nm & 310nm\\ \hline
NA&0.35~~~~~~0.5	&0.35~~~~~~0.5	&0.35~~~~~~0.5	&0.35~~~~~~0.5	&0.35~~~~~~0.5	 &0.35~~~~~~0.5	 &0.35~~~~~~0.5	 &0.35~~~~~~0.5	 \\
{1L}&0.686	0.693	~&0.644	~0.649	~&0.624	~0.626	~&0.627	~0.625	~&0.627	~0.627	~&0.645	~ 0.636	~ &0.679	~0.661	~&0.723	~0.699\\
{2L}&0.466	~0.474	~&0.420	~0.425	~&0.402	~0.404	~&0.409	~0.407	~&0.408	~0.407	~&0.433	~ 0.421	~ &0.477	~0.454	~&0.537	~0.505\\
{3L}&0.318	~0.325	~&0.281	~0.285	~&0.268	~0.269	~&0.277	~0.275	~&0.276	~0.273	~&0.301	~ 0.289	~ &0.346	~0.323	~&0.409	~0.375\\
{4L}&0.220	~0.226	~&0.192	~0.195	~&0.185	~0.185	~&0.194	~0.192	~&0.193	~0.190	~&0.216	~ 0.205	~ &0.257	~0.236	~&0.318	~0.285\\
{5L}&0.156	~0.160	~&0.135	~0.137	~&0.131	~0.131	~&0.140	~0.138	~&0.139	~0.135	~&0.160	~ 0.149	~ &0.196	~0.177	~&0.252	~0.222\\
{6L}&0.112	~0.116	~&0.097	~0.099	~&0.095	~0.095	~&0.104	~0.102	~&0.103	~0.099	~&0.121	~ 0.112	~ &0.153	~0.136	~&0.203	~0.175\\
{7L}&0.083	~0.085	~&0.072	~0.073	~&0.071	~0.071	~&0.079	~0.077	~&0.078	~0.075	~&0.093	~ 0.085	~ &0.121	~0.106	~&0.166	~0.141\\
{8L}&0.062	~0.064	~&0.054	~0.055	~&0.054	~0.054	~&0.061	~0.060	~&0.060	~0.057	~&0.073	~ 0.067	~ &0.098	~0.085	~&0.137	~0.116\\
{9L}&0.048	~0.049	~&0.042	~0.043	~&0.042	~0.042	~&0.048	~0.047	~&0.047	~0.045	~&0.059	~ 0.053	~ &0.080	~0.069	~&0.115	~0.096\\
{10L}&0.037	~0.038	~&0.033	~0.033	~&0.033	~0.033	~&0.039	~0.038	~&0.038	~0.036	~&0.048	~ 0.043	~ &0.066	~0.057	~&0.098	~0.081\\
\hline\hline
\end{tabular}
\end{center}
\end{sidewaystable}

\begin{sidewaystable}  %Table 2
\caption{I$_{2D}$(Si)/I$_0$(Si) of NL-MoS$_{2}$ flakes deposited on SiO$_2$/Si substrate as a function of N for $h_{SiO2}$=80nm, 90nm, 100nm, 110nm, and 280nm, 290nm, 300nm, 310nm. $\varepsilon_L$=2.34 eV and N.A.=0.35 and 0.50. $\tilde{n}$ of NL-MoS$_{2}$, SiO$_2$ and Si are 4.85-1.20$\bf\emph{i}$, 1.460, 4.143-0.054$\bf\emph{i}$.}
\begin{center}
\begin{tabular}{c|c|c|c|c|c|c|c|c}
  \hline\hline
  % after \\: \hline or \cline{col1-col2} \cline{col3-col4} ...
  {$h_{SiO2}$} & 80nm & 90nm & 100nm & 110nm & 280nm & 290nm & 300nm & 310nm\\ \hline
NA&0.35~~~~~~0.5	~&0.35~~~~~~0.5	~&0.35~~~~~~0.5	~&0.35~~~~~~0.5	~&0.35~~~~~~0.5	~ &0.35~~~~~~0.5	 ~ &0.35~~~~~~0.5	~ &0.35~~~~~~0.5	~ \\
{1L}&0.855	~0.862	~&0.806	~0.813	~&0.770	~0.775	~&0.748	~0.751	~&0.796	~0.817	~&0.764	~ 0.779	~ &0.746	~0.755	~&0.745	~0.746\\
{2L}&0.708	~0.719	~&0.637	~0.646	~&0.587	~0.593	~&0.561	~0.564	~&0.622	~0.652	~&0.580	~ 0.600	~ &0.560	~0.570	~&0.561	~0.560\\
{3L}&0.574	~0.587	~&0.497	~0.507	~&0.447	~0.453	~&0.423	~0.426	~&0.482	~0.514	~&0.440	~ 0.461	~ &0.423	~0.432	~&0.428	~0.426\\
{4L}&0.459	~0.472	~&0.386	~0.395	~&0.342	~0.348	~&0.323	~0.325	~&0.373	~0.403	~&0.337	~ 0.354	~ &0.324	~0.330	~&0.331	~0.327\\
{5L}&0.365	~0.377	~&0.300	~0.308	~&0.263	~0.268	~&0.249	~0.251	~&0.290	~0.315	~&0.260	~ 0.274	~ &0.250	~0.255	~&0.259	~0.255\\
{6L}&0.290	~0.300	~&0.235	~0.241	~&0.205	~0.209	~&0.194	~0.196	~&0.226	~0.248	~&0.202	~ 0.214	~ &0.196	~0.200	~&0.206	~0.201\\
{7L}&0.230	~0.239	~&0.185	~0.190	~&0.161	~0.164	~&0.154	~0.155	~&0.178	~0.196	~&0.159	~ 0.168	~ &0.156	~0.158	~&0.165	~0.161\\
{8L}&0.184	~0.191	~&0.147	~0.151	~&0.128	~0.130	~&0.123	~0.124	~&0.141	~0.156	~&0.127	~ 0.134	~ &0.125	~0.126	~&0.134	~0.130\\
{9L}&0.148	~0.154	~&0.117	~0.121	~&0.103	~0.105	~&0.100	~0.100	~&0.113	~0.125	~&0.102	~ 0.108	~ &0.102	~0.102	~&0.111	~0.106\\
{10L}&0.120	~0.125	~&0.095	~0.098	~&0.083	~0.085	~&0.082	~0.082	~&0.092	~0.101	~&0.083	~ 0.088	~ &0.083	~0.084	~&0.092	~0.088\\
\hline\hline
\end{tabular}
\end{center}
\end{sidewaystable}

\begin{sidewaystable}  %Table 3
\caption{I$_{2D}$(Si)/I$_0$(Si) of NL-WS$_{2}$ flakes deposited on SiO$_2$/Si substrate as a function of N for $h_{SiO2}$=80nm, 90nm, 100nm, 110nm, and 280nm, 290nm, 300nm, 310nm. $\varepsilon_L$=2.54 eV and N.A.=0.35 and 0.50. $\tilde{n}$ of NL-WS$_{2}$, SiO$_2$ and Si are 4.40-1.10$\bf\emph{i}$, 1.462, 4.360-0.086$\bf\emph{i}$.}
\begin{center}
\begin{tabular}{c|c|c|c|c|c|c|c|c}
  \hline\hline
  % after \\: \hline or \cline{col1-col2} \cline{col3-col4} ...
  {$h_{SiO2}$} & 80nm & 90nm & 100nm & 110nm & 280nm & 290nm & 300nm & 310nm\\ \hline
NA&0.35~~~~~~0.5	~&0.35~~~~~~0.5	~&0.35~~~~~~0.5	~&0.35~~~~~~0.5	~&0.35~~~~~~0.5	~ &0.35~~~~~~0.5	 ~ &0.35~~~~~~0.5	~ &0.35~~~~~~0.5	~ \\
{1L}&0.833	~0.839	~&0.791	~0.796	~&0.766	~0.769	~&0.760	~0.760	~&0.762	~0.766	~&0.769	~ 0.765	~ &0.790	~0.779	~&0.820	~0.804\\
{2L}&0.679	~0.689	~&0.620	~0.627	~&0.589	~0.592	~&0.583	~0.583	~&0.585	~0.589	~&0.598	~ 0.591	~ &0.631	~0.614	~&0.680	~0.654\\
{3L}&0.547	~0.558	~&0.485	~0.492	~&0.455	~0.458	~&0.453	~0.452	~&0.453	~0.456	~&0.471	~ 0.462	~ &0.511	~0.490	~&0.569	~0.538\\
{4L}&0.438	~0.448	~&0.380	~0.386	~&0.354	~0.357	~&0.355	~0.354	~&0.355	~0.357	~&0.376	~ 0.366	~ &0.418	~0.396	~&0.480	~0.447\\
{5L}&0.349	~0.359	~&0.299	~0.305	~&0.278	~0.280	~&0.282	~0.280	~&0.282	~0.282	~&0.303	~ 0.293	~ &0.345	~0.324	~&0.409	~0.375\\
{6L}&0.279	~0.288	~&0.237	~0.242	~&0.221	~0.223	~&0.227	~0.225	~&0.226	~0.225	~&0.247	~ 0.237	~ &0.288	~0.267	~&0.351	~0.318\\
{7L}&0.224	~0.232	~&0.189	~0.194	~&0.178	~0.179	~&0.184	~0.182	~&0.183	~0.182	~&0.204	~ 0.194	~ &0.243	~0.223	~&0.304	~0.271\\
{8L}&0.181	~0.188	~&0.153	~0.156	~&0.144	~0.145	~&0.151	~0.149	~&0.150	~0.148	~&0.170	~ 0.160	~ &0.207	~0.188	~&0.265	~0.234\\
{9L}&0.148	~0.153	~&0.124	~0.127	~&0.118	~0.119	~&0.125	~0.124	~&0.124	~0.122	~&0.143	~ 0.134	~ &0.178	~0.160	~&0.233	~0.203\\
{10L}&0.121	~0.125	~&0.102	~0.104	~&0.098	~0.098	~&0.105	~0.104	~&0.104	~0.102	~&0.122	~ 0.113	~ &0.154	~0.137	~&0.206	~0.178\\
\hline\hline
\end{tabular}
\end{center}
\end{sidewaystable}

\begin{sidewaystable}  %Table 4
\caption{I$_{2D}$(Si)/I$_0$(Si) of NL-WS$_{2}$ flakes deposited on SiO$_2$/Si substrate as a function of N for $h_{SiO2}$=80nm, 90nm, 100nm, 110nm, and 280nm, 290nm, 300nm, 310nm. $\varepsilon_L$=2.34 eV and N.A.=0.35 and 0.50. $\tilde{n}$ of NL-WS$_{2}$, SiO$_2$ and Si are 4.62-0.48$\bf\emph{i}$, 1.460, 4.143-0.054$\bf\emph{i}$.}
\begin{center}
\begin{tabular}{c|c|c|c|c|c|c|c|c}
  \hline\hline
  % after \\: \hline or \cline{col1-col2} \cline{col3-col4} ...
  {$h_{SiO2}$} & 80nm & 90nm & 100nm & 110nm & 280nm & 290nm & 300nm & 310nm\\ \hline
NA&0.35~~~~~~0.5	~&0.35~~~~~~0.5	~&0.35~~~~~~0.5	~&0.35~~~~~~0.5	~&0.35~~~~~~0.5	~ &0.35~~~~~~0.5	 ~ &0.35~~~~~~0.5	~ &0.35~~~~~~0.5	~ \\
{1L}&0.970	~0.977	~&0.920	~0.927	~&0.875	~0.882	~&0.844	~0.848	~&0.907	~0.930	~&0.866	~ 0.886	~ &0.839	~0.853	~&0.826	~0.833\\
{2L}&0.908	~0.922	~&0.819	~0.831	~&0.747	~0.757	~&0.701	~0.707	~&0.798	~0.837	~&0.734	~ 0.765	~ &0.695	~0.715	~&0.679	~0.688\\
{3L}&0.824	~0.842	~&0.710	~0.725	~&0.626	~0.638	~&0.576	~0.583	~&0.686	~0.734	~&0.612	~ 0.647	~ &0.571	~0.592	~&0.557	~0.565\\
{4L}&0.728	~0.748	~&0.603	~0.619	~&0.519	~0.530	~&0.471	~0.478	~&0.578	~0.629	~&0.505	~ 0.540	~ &0.467	~0.487	~&0.458	~0.463\\
{5L}&0.629	~0.650	~&0.505	~0.521	~&0.426	~0.437	~&0.385	~0.391	~&0.482	~0.532	~&0.415	~ 0.447	~ &0.382	~0.399	~&0.377	~0.380\\
{6L}&0.535	~0.555	~&0.419	~0.433	~&0.349	~0.359	~&0.315	~0.320	~&0.399	~0.444	~&0.340	~ 0.368	~ &0.313	~0.327	~&0.312	~0.313\\
{7L}&0.450	~0.468	~&0.346	~0.359	~&0.286	~0.294	~&0.259	~0.263	~&0.328	~0.369	~&0.279	~ 0.303	~ &0.258	~0.269	~&0.259	~0.259\\
{8L}&0.376	~0.392	~&0.285	~0.296	~&0.235	~0.242	~&0.213	~0.217	~&0.271	~0.306	~&0.229	~ 0.249	~ &0.214	~0.222	~&0.217	~0.216\\
{9L}&0.313	~0.327	~&0.235	~0.245	~&0.194	~0.200	~&0.177	~0.180	~&0.223	~0.253	~&0.190	~ 0.206	~ &0.178	~0.184	~&0.183	~0.181\\
{10L}&0.260	~0.273	~&0.195	~0.203	~&0.161	~0.166	~&0.148	~0.150	~&0.185	~0.210	~&0.158	~ 0.171	~ &0.149	~0.154	~&0.155	~0.153\\
\hline\hline
\end{tabular}
\end{center}
\end{sidewaystable}

\begin{sidewaystable}  %Table 5
\caption{I$_{2D}$(Si)/I$_0$(Si) of NL-WeS$_{2}$ flakes deposited on SiO$_2$/Si substrate as a function of N for $h_{SiO2}$=80nm, 90nm, 100nm, 110nm, and 280nm, 290nm, 300nm, 310nm. $\varepsilon_L$=2.54 eV and N.A.=0.35 and 0.50. $\tilde{n}$ of NL-WeS$_{2}$, SiO$_2$ and Si are 4.22-1.86$\bf\emph{i}$, 1.462, 4.360-0.086$\bf\emph{i}$.}
\begin{center}
\begin{tabular}{c|c|c|c|c|c|c|c|c}
  \hline\hline
  % after \\: \hline or \cline{col1-col2} \cline{col3-col4} ...
  {$h_{SiO2}$} & 80nm & 90nm & 100nm & 110nm & 280nm & 290nm & 300nm & 310nm\\ \hline
NA&0.35~~~~~~0.5	~&0.35~~~~~~0.5	~&0.35~~~~~~0.5	~&0.35~~~~~~0.5	~&0.35~~~~~~0.5	~ &0.35~~~~~~0.5	 ~ &0.35~~~~~~0.5	~ &0.35~~~~~~0.5	~ \\
{1L}&0.728	~0.733	~&0.700	~0.703	~&0.690	~0.690	~&0.698	~0.695	~&0.697	~0.694	~&0.717	~ 0.707	~ &0.748	~0.732	~&0.786	~0.766\\
{2L}&0.533	~0.539	~&0.498	~0.502	~&0.489	~0.489	~&0.501	~0.498	~&0.499	~0.495	~&0.528	~ 0.513	~ &0.572	~0.549	~&0.628	~0.598\\
{3L}&0.394	~0.400	~&0.362	~0.365	~&0.354	~0.355	~&0.369	~0.365	~&0.367	~0.362	~&0.398	~ 0.382	~ &0.446	~0.421	~&0.510	~0.476\\
{4L}&0.295	~0.300	~&0.268	~0.271	~&0.263	~0.263	~&0.278	~0.275	~&0.276	~0.270	~&0.306	~ 0.291	~ &0.354	~0.329	~&0.419	~0.384\\
{5L}&0.223	~0.228	~&0.202	~0.204	~&0.199	~0.199	~&0.214	~0.210	~&0.211	~0.206	~&0.240	~ 0.226	~ &0.285	~0.262	~&0.349	~0.315\\
{6L}&0.172	~0.176	~&0.155	~0.157	~&0.154	~0.153	~&0.167	~0.164	~&0.165	~0.160	~&0.191	~ 0.178	~ &0.233	~0.211	~&0.294	~0.261\\
{7L}&0.134	~0.137	~&0.120	~0.122	~&0.121	~0.120	~&0.133	~0.130	~&0.131	~0.126	~&0.154	~ 0.143	~ &0.193	~0.173	~&0.250	~0.219\\
{8L}&0.106	~0.109	~&0.095	~0.096	~&0.096	~0.096	~&0.107	~0.105	~&0.105	~0.101	~&0.126	~ 0.116	~ &0.162	~0.143	~&0.215	~0.186\\
{9L}&0.085	~0.087	~&0.076	~0.077	~&0.078	~0.077	~&0.088	~0.086	~&0.086	~0.082	~&0.105	~ 0.096	~ &0.137	~0.120	~&0.186	~0.159\\
{10L}&0.068	~0.070	~&0.062	~0.063	~&0.063	~0.063	~&0.073	~0.071	~&0.071	~0.068	~&0.088	~ 0.080	~ &0.117	~0.102	~&0.162	~0.138\\
\hline\hline
\end{tabular}
\end{center}
\end{sidewaystable}

	~	~	~	~	~	~
\begin{sidewaystable}  %Table 6
\caption{I$_{2D}$(Si)/I$_0$(Si) of NL-WeS$_{2}$ flakes deposited on SiO$_2$/Si substrate as a function of N for $h_{SiO2}$=80nm, 90nm, 100nm, 110nm, and 280nm, 290nm, 300nm, 310nm. $\varepsilon_L$=2.34 eV and N.A.=0.35 and 0.50. $\tilde{n}$ of NL-WeS$_{2}$, SiO$_2$ and Si are 4.64-1.40$\bf\emph{i}$, 1.460, 4.143-0.054$\bf\emph{i}$.}
\begin{center}
\begin{tabular}{c|c|c|c|c|c|c|c|c}
  \hline\hline
  % after \\: \hline or \cline{col1-col2} \cline{col3-col4} ...
  {$h_{SiO2}$} & 80nm & 90nm & 100nm & 110nm & 280nm & 290nm & 300nm & 310nm\\ \hline
NA&0.35~~~~~~0.5	~&0.35~~~~~~0.5	~&0.35~~~~~~0.5	~&0.35~~~~~~0.5	~&0.35~~~~~~0.5	~ &0.35~~~~~~0.5	 ~ &0.35~~~~~~0.5	~ &0.35~~~~~~0.5	~ \\
{1L}&0.822	~0.829	~&0.778	~0.784	~&0.746	~0.750	~&0.729	~0.730	~&0.769	~0.788	~&0.741	~ 0.754	~ &0.728	~0.735	~&0.731	~0.729\\
{2L}&0.660	~0.670	~&0.598	~0.606	~&0.556	~0.561	~&0.536	~0.538	~&0.586	~0.612	~&0.550	~ 0.567	~ &0.537	~0.544	~&0.543	~0.540\\
{3L}&0.523	~0.534	~&0.457	~0.466	~&0.417	~0.422	~&0.400	~0.401	~&0.446	~0.472	~&0.412	~ 0.428	~ &0.401	~0.407	~&0.410	~0.406\\
{4L}&0.412	~0.422	~&0.351	~0.358	~&0.315	~0.320	~&0.302	~0.304	~&0.341	~0.365	~&0.312	~ 0.326	~ &0.304	~0.308	~&0.315	~0.310\\
{5L}&0.323	~0.333	~&0.270	~0.277	~&0.241	~0.245	~&0.232	~0.233	~&0.262	~0.283	~&0.239	~ 0.250	~ &0.234	~0.237	~&0.246	~0.240\\
{6L}&0.255	~0.263	~&0.210	~0.216	~&0.187	~0.190	~&0.181	~0.181	~&0.204	~0.221	~&0.185	~ 0.194	~ &0.183	~0.185	~&0.195	~0.189\\
{7L}&0.202	~0.209	~&0.165	~0.170	~&0.147	~0.149	~&0.143	~0.143	~&0.160	~0.174	~&0.146	~ 0.153	~ &0.145	~0.146	~&0.156	~0.151\\
{8L}&0.161	~0.167	~&0.131	~0.135	~&0.116	~0.118	~&0.114	~0.114	~&0.127	~0.139	~&0.116	~ 0.122	~ &0.116	~0.117	~&0.127	~0.122\\
{9L}&0.129	~0.134	~&0.105	~0.108	~&0.094	~0.095	~&0.092	~0.092	~&0.102	~0.111	~&0.093	~ 0.098	~ &0.095	~0.094	~&0.105	~0.100\\
{10L}&0.105	~0.109	~&0.085	~0.088	~&0.076	~0.077	~&0.076	~0.076	~&0.082	~0.090	~&0.076	~ 0.080	~ &0.078	~0.077	~&0.087	~0.082\\
\hline\hline
\end{tabular}
\end{center}
\end{sidewaystable}

\begin{sidewaystable}  %Table 7
\caption{I$_{2D}$(Si)/I$_0$(Si) of NLG flakes deposited on SiO$_2$/Si substrate as a function of N for $h_{SiO2}$=80nm, 90nm, 100nm, 110nm, and 280nm, 290nm, 300nm, 310nm. $\varepsilon_L$=2.34 eV and N.A.=0.35 and 0.50. $\tilde{n}$ of NLG, SiO$_2$ and Si are 2.725-1.366$\bf\emph{i}$, 1.460, 4.143-0.054$\bf\emph{i}$.}
\begin{center}
\begin{tabular}{c|c|c|c|c|c|c|c|c}
  \hline\hline
  % after \\: \hline or \cline{col1-col2} \cline{col3-col4} ...
  {$h_{SiO2}$} & 80nm & 90nm & 100nm & 110nm & 280nm & 290nm & 300nm & 310nm\\ \hline
NA&0.35~~~~~~0.5	~&0.35~~~~~~0.5	~&0.35~~~~~~0.5	~&0.35~~~~~~0.5	~&0.35~~~~~~0.5	~ &0.35~~~~~~0.5	 ~ &0.35~~~~~~0.5	~ &0.35~~~~~~0.5	~ \\
{1L}&0.938	~0.939	~&0.929	~0.930	&0.924	~0.924	&0.923	~0.923	&0.927	~0.931	&0.924	~ 0.926	~ &0.924	~0.924	&0.928	~0.926\\
{2L}&0.879	~0.882	&0.863	~0.866	&0.854	~0.856	&0.853	~0.853	&0.861	~0.868	&0.854	~ 0.858	~ &0.855	~0.855	&0.863	~0.859\\
{3L}&0.824	~0.828	&0.803	~0.806	&0.791	~0.793	&0.790	~0.790	&0.800	~0.809	&0.791	~ 0.796	~ &0.792	~0.792	&0.803	~0.798\\
{4L}&0.773	~0.778	&0.748	~0.751	&0.733	~0.735	&0.732	~0.732	&0.744	~0.755	&0.733	~ 0.739	~ &0.735	~0.735	&0.749	~0.742\\
{5L}&0.726	~0.731	&0.696	~0.701	&0.681	~0.683	&0.680	~0.679	&0.692	~0.704	&0.680	~ 0.687	~ &0.683	~0.682	&0.699	~0.691\\
{6L}&0.681	~0.687	&0.649	~0.654	&0.632	~0.635	&0.632	~0.631	&0.644	~0.658	&0.632	~ 0.639	~ &0.636	~0.635	&0.653	~0.645\\
{7L}&0.639	~0.646	&0.606	~0.611	&0.588	~0.591	&0.588	~0.588	&0.601	~0.615	&0.588	~ 0.595	~ &0.592	~0.591	&0.611	~0.602\\
{8L}&0.600	~0.607	&0.565	~0.571	&0.548	~0.551	&0.548	~0.548	&0.561	~0.575	&0.548	~ 0.555	~ &0.553	~0.551	&0.573	~0.563\\
{9L}&0.564	~0.571	&0.528	~0.534	&0.511	~0.513	&0.511	~0.511	&0.523	~0.539	&0.511	~ 0.518	~ &0.516	~0.515	&0.538	~0.527\\
{10L}&0.530	~0.538	&0.494	~0.500	&0.477	~0.479	&0.478	~0.477	&0.489	~0.504	&0.477	~ 0.484	~ &0.483	~0.481	&0.505	~0.494\\
{11L}&0.499	~0.506	&0.463	~0.468	&0.445	~0.448	&0.447	~0.446	&0.458	~0.473	&0.446	~ 0.453	~ &0.452	~0.450	&0.475	~0.464\\
{12L}&0.469	~0.477	&0.433	~0.439	&0.417	~0.419	&0.418	~0.418	&0.429	~0.444	&0.417	~ 0.424	~ &0.424	~0.422	&0.447	~0.436\\
{13L}&0.442	~0.449	&0.406	~0.412	&0.390	~0.393	&0.392	~0.392	&0.402	~0.417	&0.391	~ 0.397	~ &0.398	~0.395	&0.421	~0.410\\
{14L}&0.416	~0.423	&0.381	~0.386	&0.366	~0.368	&0.368	~0.368	&0.377	~0.391	&0.367	~ 0.373	~ &0.374	~0.371	&0.398	~0.386\\
{15L}&0.392	~0.399	&0.358	~0.363	&0.343	~0.345	&0.346	~0.346	&0.354	~0.368	&0.344	~ 0.350	~ &0.352	~0.349	&0.376	~0.364\\
\hline\hline
\end{tabular}
\end{center}
\end{sidewaystable}

\begin{sidewaystable} %Table 8
\caption{I$_{2D}$(Si)/I$_0$(Si) of NLG flakes deposited on SiO$_2$/Si substrate as a function of N for $h_{SiO_2}$=80nm, 90nm, 100nm, 110nm, and 280nm, 290nm, 300nm, 310nm. $\varepsilon_L$=1.96 eV and N.A.=0.35 and 0.50. $\tilde{n}$ of NLG, SiO$_2$ and Si are 2.819-1.450$\bf\emph{i}$, 1.457, 3.879-0.021$\bf\emph{i}$.}
\begin{center}
\begin{tabular}{c|c|c|c|c|c|c|c|c}
  \hline\hline
  % after \\: \hline or \cline{col1-col2} \cline{col3-col4} ...
  {$h_{SiO2}$} & 80nm & 90nm & 100nm & 110nm & 280nm & 290nm & 300nm & 310nm\\ \hline
NA&0.35~~~~~~0.5	&0.35~~~~~~0.5	&0.35~~~~~~0.5	&0.35~~~~~~0.5	&0.35~~~~~~0.5	~ &0.35~~~~~~0.5	 ~ &0.35~~~~~~0.5	~ &0.35~~~~~~0.5	~ \\
{1L}&0.955	~0.957	&0.948	~0.949	&0.940	~0.942	&0.935	~0.936	&0.972	~0.975	&0.966	~ 0.969	~ &0.959	~0.963	&0.951	~0.955\\
{2L}&0.913	~0.915	&0.898	~0.900	&0.885	~0.887	&0.874	~0.876	&0.945	~0.950	&0.933	~ 0.939	~ &0.919	~0.926	&0.904	~0.912\\
{3L}&0.872	~0.875	&0.851	~0.854	&0.832	~0.836	&0.818	~0.821	&0.918	~0.925	&0.901	~ 0.910	~ &0.881	~0.891	&0.860	~0.871\\
{4L}&0.832	~0.836	&0.806	~0.811	&0.783	~0.788	&0.766	~0.770	&0.892	~0.901	&0.869	~ 0.881	~ &0.844	~0.857	&0.818	~0.832\\
{5L}&0.794	~0.799	&0.764	~0.769	&0.738	~0.742	&0.718	~0.722	&0.866	~0.877	&0.838	~ 0.852	~ &0.808	~0.824	&0.777	~0.794\\
{6L}&0.758	~0.764	&0.724	~0.730	&0.695	~0.700	&0.674	~0.678	&0.840	~0.853	&0.808	~ 0.824	~ &0.773	~0.792	&0.739	~0.758\\
{7L}&0.723	~0.730	&0.686	~0.692	&0.655	~0.661	&0.632	~0.637	&0.815	~0.830	&0.779	~ 0.797	~ &0.740	~0.761	&0.702	~0.723\\
{8L}&0.690	~0.697	&0.650	~0.657	&0.617	~0.623	&0.594	~0.598	&0.790	~0.807	&0.751	~ 0.771	~ &0.709	~0.731	&0.668	~0.690\\
{9L}&0.658	~0.666	&0.617	~0.624	&0.582	~0.589	&0.558	~0.563	&0.766	~0.784	&0.723	~ 0.745	~ &0.678	~0.702	&0.635	~0.658\\
{10L}&0.628	~0.636	&0.585	~0.592	&0.550	~0.556	&0.525	~0.530	&0.743	~0.762	&0.697	~ 0.720	~ &0.649	~0.674	&0.604	~0.628\\
{11L}&0.599	~0.607	&0.555	~0.562	&0.519	~0.525	&0.494	~0.499	&0.720	~0.740	&0.671	~ 0.696	~ &0.621	~0.647	&0.574	~0.600\\
{12L}&0.572	~0.580	&0.526	~0.534	&0.490	~0.497	&0.466	~0.470	&0.697	~0.719	&0.646	~ 0.672	~ &0.594	~0.621	&0.546	~0.572\\
{13L}&0.546	~0.554	&0.500	~0.507	&0.463	~0.470	&0.439	~0.444	&0.675	~0.698	&0.622	~ 0.649	~ &0.569	~0.596	&0.520	~0.546\\
{14L}&0.521	~0.529	&0.474	~0.482	&0.438	~0.445	&0.414	~0.419	&0.654	~0.678	&0.599	~ 0.626	~ &0.544	~0.573	&0.495	~0.521\\
{15L}&0.497	~0.506	&0.450	~0.458	&0.414	~0.421	&0.391	~0.395	&0.633	~0.658	&0.576	~ 0.605	~ &0.521	~0.550	&0.471	~0.498\\
\hline\hline
\end{tabular}
\end{center}
\end{sidewaystable}